\documentclass[]{mn2e}
\usepackage{pstricks}
\usepackage{natbib}
\usepackage{amssymb}
\usepackage{amsmath}
\usepackage{graphicx}
\newcommand{\spcfs}{$\xi_s(\sigma,\pi)$}

\newcommand{\Qcfr}{$\xi_{r,QCF}$}

\begin{document}

\title[Determining the growth rate from simulations]{Determining accurate measurements of the growth rate from the galaxy correlation function in simulations}
\author[Contreras et al.]{\parbox[t]{\textwidth}{Carlos Contreras$^{1,2}$, Chris Blake$^1$, Gregory B.\ Poole$^1$, Felipe Marin$^1$}
\\ \\ $^1$ Centre for Astrophysics \& Supercomputing, Swinburne University of Technology, P.O. Box 218, Hawthorn, VIC 3122, Australia,
\\ \\ $^2$ Las Campanas Observatory, Casilla 601, La Serena, Chile}

\maketitle

\begin{abstract}
  We use high-resolution N-body simulations to develop a new,
  flexible, empirical approach for measuring the growth rate from
  redshift-space distortions (RSD) in the 2-point galaxy correlation
  function.  We quantify the systematic error in measuring the growth
  rate in a $1 \, h^{-3}$ Gpc$^3$ volume over a range of redshifts,
  from the dark matter particle distribution and a range of halo-mass
  catalogues with a number density comparable to the latest
  large-volume galaxy surveys such as the WiggleZ Dark Energy Survey
  and the Baryon Oscillation Spectroscopic Survey (BOSS).  Our
  simulations allow us to span halo masses with bias factors ranging
  from unity (probed by emission-line galaxies) to more massive haloes
  hosting Luminous Red Galaxies.  We show that the measured growth
  rate is sensitive to the model adopted for the small-scale
  real-space correlation function, and in particular that the
  ``standard'' assumption of a power-law correlation function can
  result in a significant systematic error in the growth rate
  determination.  We introduce a new, empirical fitting function that
  produces results with a lower (5-10\%) amplitude of systematic
  error.  We also introduce a new technique which permits the galaxy
  pairwise velocity distribution, the quantity which drives the
  non-linear growth of structure, to be measured as a non-parametric
  stepwise function.  Our (model-independent) results agree well with
  an exponential pairwise velocity distribution, expected from
  theoretical considerations, and are consistent with direct
  measurements of halo velocity differences from the parent
  catalogues.  In a companion paper we present the application of our
  new methodology to the WiggleZ Survey dataset.
\end{abstract}

\section{Introduction}

The growth rate of cosmic structure is a key parameter which
quantifies the cosmological model.  While distance-redshift probes
such as Type Ia supernovae \citep[e.g.][]{Rs98,Pm99,Kow08,Hi09,Am10}
or baryon acoustic oscillations
\citep[e.g.][]{Blk03,Seo03,Ei05,Pc10,Blk11} give us information about
the cosmic expansion history, and observations of the Cosmic Microwave
Background radiation allow the study of the composition and physics of
the early Universe \citep[e.g.][]{Ko10}, the growth rate describes how
the small density perturbations present in the early Universe evolve
to form the large-scale structure which populates the Universe today.
Hence the growth rate provides a fundamental test of the laws of
gravity which operate in the expanding cosmos.

Measurements of the growth rate have assumed a special importance as
evidence has accumulated that the expansion of the Universe has
entered a phase of acceleration.  Two principle explanations have been
put forward for accelerating expansion.  The first is the presence of
some unknown material constituent of the Universe with the exotic
property of negative pressure, whose energy density has become
dominant within the last half of the age of the Universe.  This
material is known as ``Dark Energy''.  The second explanation is that
our current theory of gravity, General Relativity, must be modified on
large scales to account for the observations without invoking exotic
constituents.  Various methods for modifying General Relativity have
been explored \citep[see][]{Tsu10}. Measurements of the growth rate of
structure over different cosmic epochs can help discriminate between
these two physical interpretations of the observations
\citep{LJ03,LC07,G08,W08}.

The growth rate of structure within a given cosmological model can be
derived by solving the differential equation for the linear density of
matter perturbations $\delta$ at scale factor $a$ in an expanding
Universe \citep{Pb80}.  The growth rate at redshift $z$ is defined by
$f(z) = d\ln{\delta}/d\ln{a}$ and is well-approximated in a variety of
dark energy models by $f(z) = \Omega_m(z)^\gamma$, where $\Omega_m(z)$
is the matter density relative to the critical density, and $\gamma$
is a phenomenological parameter which takes the value $0.55$ for
General Relativity \citep{LC07}.

A powerful method for measuring the growth rate is to exploit the
anisotropic signature it imprints in the clustering within galaxy
redshift surveys, known as redshift-space distortions \citep[RSD, see
  e.g.][]{Hk03,Tm04,Tm06,dAn08,G08,Ok08,CG09,Blk11} .  The growth of
structure is driven by coherent flows of matter into clusters and
superclusters, and the resulting coherent galaxy peculiar velocities
produce a correlated signature in the pattern of 2-point galaxy
clustering.  The galaxy clustering anisotropy on large scales is
parameterized by a measured distortion parameter $\beta$, which is
related to the growth rate by $f = b \, \beta$, where $b$ is the bias
parameter which relates galaxy overdensity to matter overdensity.

An alternative method to map out the clustering pattern and measure
the growth rate is weak gravitational lensing, which involves
modelling the observed correlated alignment of distant background
galaxy shapes by the foreground large-scale structure.  However, the
current sensitivity of the method is limited \cite[e.g.][]{Be09},
despite extremely promising future prospects.  Other methods that have
been developed to constrain the growth rate include the luminosity
function and gas-mass fraction of X-ray selected clusters
\citep{Ra09}, and galaxy bulk flows measured in the local neighborhood
\citep{AL08,Wk09,Nu11}.  However, at the present time redshift-space
distortions in galaxy clustering represent the most accurate method
for measuring the growth rate of structure.

Utilizing redshift-space distortions as a cosmological probe has two
requirements.  Firstly, a large galaxy redshift survey must be
performed in order to reduce the sources of error in the measurement,
cosmic variance and galaxy shot noise.  Secondly, reliable models must
be fitted to the clustering measurements.  The second issue,
correlation-function modelling, is the subject of the current paper,
whereas applications of these techniques to the WiggleZ Dark Energy
Survey \citep{Dw10} are presented in a companion paper.

In order to model the two-dimensional redshift-space galaxy
correlation function transverse and parallel to the line-of-sight,
\spcfs, we must model the pattern of peculiar velocities acquired by
galaxies as part of the growth of structure.  There are (crudely
speaking) two sources for these peculiar velocities: the random
velocity a galaxy possesses with respect to its own group or cluster,
and the velocity the galaxy acquires as part of bulk flows into bigger
structures such as superclusters.  These two effects distort the shape
of the measured \spcfs\ in a characteristic way which can be modelled
and split from the underlying isotropic real-space correlation
function.  The challenge is to measure the redshift-space distortion
parameter due to bulk flows, $\beta$, in a manner free from systematic
error, and relate that to the growth rate of structure using parallel
measurements of the galaxy bias factor.

The aim of this paper is to address the obstacles which must be
overcome to obtain reliable measurements of the cosmic growth rate
from modelling the correlation function, namely: (1) provision of an
accurate model for the underlying isotropic real-space correlation
function, (2) modelling the non-linear and random effects of galaxy
pairwise velocities, (3) determining an accurate measurement of the
galaxy bias factor.  We assume throughout that the background
cosmological parameters are known, for example from Cosmic Microwave
Background observations, hence we neglect Alcock-Paczynski distortions
in this study.

There currently exists no model which is able to describe accurately
the matter distribution and its clustering properties over the
complete range of scales of our interest: $0 - 50 \, h^{-1}$ Mpc.  At
least three different physical regimes can be identified which shape
the galaxy correlation function and redshift-space distortions in this
range: the non-linear, quasi-linear and linear regimes.  On large
scales ($\gtrsim 20 \, h^{-1}$ Mpc), perturbation theory provides a
good description of the growth of the clustering pattern
\citep{Bnd02,CS06,Ni09,Cs09,Ta09}, a continuity equation holds between
velocity and density, and the redshift-space distortion pattern is
well-described by a simple anisotropy known as the ``Kaiser limit''
\citep{Ka87}.  At intermediate scales ($\lesssim 20 \, h^{-1}$ Mpc),
in the quasi-linear regime, perturbation theory breaks down
\citep{Pb80,HC98,Rc98,L02,Sc04} and significant corrections are needed
to the Kaiser limit formulation.  At small scales ($\lesssim 3 \,
h^{-1}$ Mpc), in the non-linear or ``1-halo'' regime, the matter
distribution is virialized in groups and clusters of galaxies and high
velocity dispersions imprint the ``fingers-of-god'' feature into the
correlation function.  All of these physical ingredients leave
different signatures in the correlation function, currently making it
difficult to construct a physically-motivated model which is valid
across the entire range of scales.  In this study we instead focus on
empirical approaches for modelling the observations.

N-body dark matter simulations, and the halo catalogues that can be
constructed from them, are a powerful method for modelling the full
range of non-linear processes described above.  Because the
cosmological parameters and input growth rate in the simulations are
known, we can study the systematic errors that arise from using
particular algorithms to extract these observables.  Simulations allow
us to see in detail pieces of information that are typically hidden in
the real data samples, such as the velocity distributions in
virialized clusters or the shape of the real-space correlation
function.

Other studies in the literature have used simulations to test and
improve models fitted to RSD.  In \citet{CG09} the pairwise velocity
distribution in redshift-space was studied in detail and found to be
scale-dependent, leading to the inclusion in the model of two
independent velocity dispersion terms applying at scales smaller and
larger than $2 \, h^{-1}$ Mpc \citep[see also][]{Sl06}.  \citet{Jn11}
used N-body simulations to test models of varying complexity for
recovering the true value of RSD for different cosmologies via the
power spectrum.  They demonstrate that linear models by themselves do
not extract an unbiased growth rate \citep[see
  also][]{Ma08,Ta09,Kw11,Ok11,Bi12}.

In our work we use a new set of N-body simulations – the Gigaparsec
WiggleZ (GiggleZ) Simulations (Poole et al., in prep) -- to develop a
new, empirical method for extracting the growth rate of structure from
a range of halo catalogues and redshifts.  The distinguishing feature
of these simulations is their low particle mass in comparison with
most large-volume simulations, which is appropriate for modelling the
relatively low-mass haloes probed by emission-line galaxies mapped by
the WiggleZ Dark Energy Survey.  In contrast to the studies cited
above, which largely considered the galaxy power spectrum, we base our
analysis on the galaxy correlation function.  This is a commonly-used
statistic for quantifying galaxy clustering which has several merits
including that (1) different physical processes (such as shot-noise)
are confined to distinct sets of scales, and (2) it is less sensitive
than the power spectrum to modelling the survey selection function.
In the absence of a complete physical model, it has been standard in
the literature to fit a power-law model as the real-space correlation
function when modelling the data in redshift-space, and include
non-linearities via an exponential pairwise velocity distribution
function.  In this paper we use the GiggleZ simulations to critically
examine these assumptions, proposing a new, improved empirical fitting
function and new techniques for studying the non-linear velocity
dispersion.

Our paper is structured as follows: in Section \ref{sec:gigglez} we
describe the N-body simulations we employ in more detail.  In Section
\ref{sec:data} we outline how the 2-point correlation function is
measured from the simulation data, and in Section \ref{sec:models} we
specify the models we fit to these measurements.  In Sections
\ref{sec:dmfit} and \ref{sec:halofit} we describe the performance and
results of these fits to the correlation functions measured from the
dark matter distribution and halo-mass catalogues, respectively.  In
Section \ref{sec:bias} we discuss the determination of the galaxy bias
parameter from these data, and in Section \ref{sec:conc} we summarize
and discuss our findings.

\section{The GiggleZ simulation}
\label{sec:gigglez}

We conducted our analysis using the Gigaparsec WiggleZ Survey
simulations (GiggleZ).  The main simulation, which we utilize here, is
a $2160^3$ particle dark matter simulation run in a periodic box $1\,
h^{-1}$ Gpc on a side.  The resulting particle mass of this simulation
is $7.5\times10^9\, h^{-1} M_\odot$ which permits us to resolve bound
systems with masses $\gtrsim 1.5\times10^{11}\, h^{-1} M_\odot$,
facilitating studies of haloes with clustering bias factors ranging
from near unity (e.g.\ galaxies in the WiggleZ Dark Energy Survey) to
in excess of 2 (e.g.\ Luminous Red Galaxies (LRGs) in the Sloan
Digital Sky Survey).  A WMAP-5 cosmology with
$(\Omega_\Lambda,\Omega_M,\Omega_b,h,\sigma_8,n)$ =
(0.727,0.273,0.0456,0.705,0.812,0.960) was assumed for this
simulation, with the initial conditions constructed to yield a CAMB
\citep{LC99} power spectrum for a starting redshift of $z=49$ using
the Zeldovich approximation \citep{Zel1970,Bu1992}.

Bound structures were identified using Subfind \citep{Sp2001},
which uses a friends-of-friends (FoF) scheme followed by a
substructure analysis to identify bound overdensities within each FoF
halo.  We use the Subfind substructures for all our analysis in this
paper and use the value of each halo's maximum circular velocity
$V_{max}$ as a proxy for mass.  This choice of mass proxy was made to
increase the robustness and reproducibility of our results, since it
avoids many numerical effects and biases associated with specific halo
finding schemes and halo mass definitions.  We use the centre of mass
velocities of each halo when computing redshift-space distortions.

\begin{figure}
\centering
\includegraphics[width=75mm]{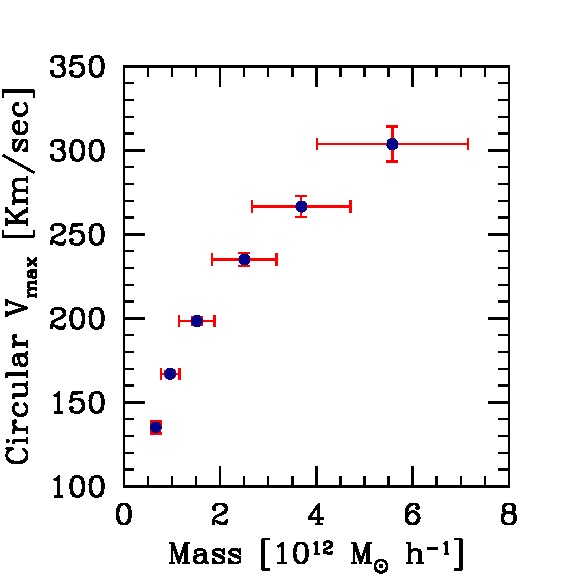}
\caption{The spread of maximum circular velocities and halo masses
  within the 6 GiggleZ simulation halo-mass catalogues analyzed in
  this study, for the $z=0$ snapshot. Each catalogue contains 250,000
  position and velocity entries in a $1\, h^{-1}$ Gpc box, generating
  a similar number density to that observed in the WiggleZ Dark Energy
  Survey or Baryon Oscillation Spectroscopic Survey.}
\label{fig:haloes}
\end{figure}

In order to explore clustering systematics as a function of halo mass,
we rank-ordered the GiggleZ substructures by their maximum circular
velocities and selected contiguous groupings of 250,000 systems.  From
this we chose a series of 6 halo groupings, ranging in halo mass from
WiggleZ galaxies to Sloan LRGs (with median values of
$V_{max}=130,160,190,230,260,300$ km/s) with a number density $2.5
\times 10^{-4} \, h^3$ Mpc$^{-3}$, which is well-matched to that of
the WiggleZ survey and ongoing Baryon Oscillation Spectroscopic
Survey.  Figure \ref{fig:haloes} illustrates the ranges of maximum
circular velocities and halo masses contained in these six catalogues
for the $z=0$ snapshot.  We note that the completeness limit of the
halo catalogue is around 120 km/s at $z=0$.

\section{Measurements of the 2D correlation function in the GiggleZ simulation}
\label{sec:data}

In this paper we quantify clustering using the two-dimensional 2-point
correlation function.  This statistic is determined by counting the
number of unique galaxy pairs as a function of transverse and parallel
separation to the line-of-sight ($\sigma$ and $\pi$), and comparing
the result to a similar pair-count performed on a randomly-distributed
catalogue.  We averaged over 30 random catalogues, each containing an
equal number of particles as the dataset.  We used two estimators to
measure the correlation function.  The first is the \cite{LS93}
estimator, the standard minimum-variance procedure performed with real
galaxy catalogues and corresponding random catalogues. In the second
estimation we exploit the fact that our simulation is a cube with
periodic boundary conditions and only count data-pairs, estimating the
random pair-count using analytic methods.  The results agree very
closely, and the second method allows a much more rapid computation.
We computed the pair counts in square $(\sigma,\pi)$ bins with side $2
\, h^{-1}$ Mpc, up to $40\, h^{-1}$ Mpc in $\sigma$ and $30\, h^{-1}$
Mpc in $\pi$ (300 data bins).

\begin{figure}
\centering
\includegraphics[width=75mm]{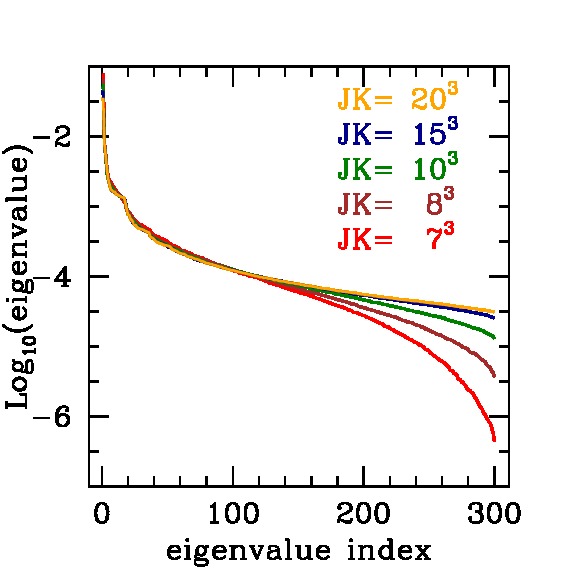}
\caption{The curve of eigenvalues in a Singular Value Decomposition of
  the covariance matrix which results as we vary the number of
  jack-knife subregions.  The robustness of the covariance matrix
  improves asymptotically until we reach $N_{JK}=20^3$, which is
  approximately the maximum number of sub-volumes in which the
  original data cube can be divided whilst still retaining dimensions
  bigger than the relevant scales in our 2-point correlation function
  measurement. The shape of these curves is similar for both methods
  of calculating the covariance matrix described in Section
  \ref{sec:data}. In the case of lower numbers of jack-knife
  subdivisions, we were forced to truncate the lowest eigenvalues of
  the covariance matrix to attain stable fits, whilst for
  $N_{JK}>10^3$ full and truncated covariances gave consistent
  results.}
\label{fig:svd}
\end{figure}

We obtained the data covariances using the jack-knife procedure, in
which we divided the $1 \, h^{-1}$ Gpc cube into a number $N_{JK}$ of
identical jack-knife regions.  Singular Value Decomposition (SVD)
analysis of the resulting covariance matrices is crucial for
understanding their robustness in the $\chi^2$ fitting procedure.
Figure \ref{fig:svd} illustrates the spectrum of SVD eigenvalues for a
range of choices of the number of jack-knife regions.  We found that
for our dataset, $7^3$ jack-knife subvolumes produced noisy covariance
matrices with both of the employed methods, and this effect is not
completely ameliorated if the lowest eigenvalues are truncated in the
re-constructed covariance matrix.  SVD analysis showed that increasing
the number of jack-knife sub-regions improves asymptotically the
quality of the covariance matrix.  In our case values in the range
$10^3$ to $20^3$ are good choices.  For our default choice $N_{JK} =
10^3$, we note that the size of each jack-knife region is $100 \,
h^{-1}$ Mpc, significantly exceeding the clustering scales of
interest.  We checked that the growth-rate measurements presented in
this paper do not depend significantly on whether we use the full
covariance matrix, a truncated matrix in which the lowest-amplitude
eigenvalues of an SVD decomposition are excluded, or a diagonal error
matrix.

\begin{figure*}
\centering
\begin{tabular}{cc}
\includegraphics[angle=-90,width=0.5\linewidth]{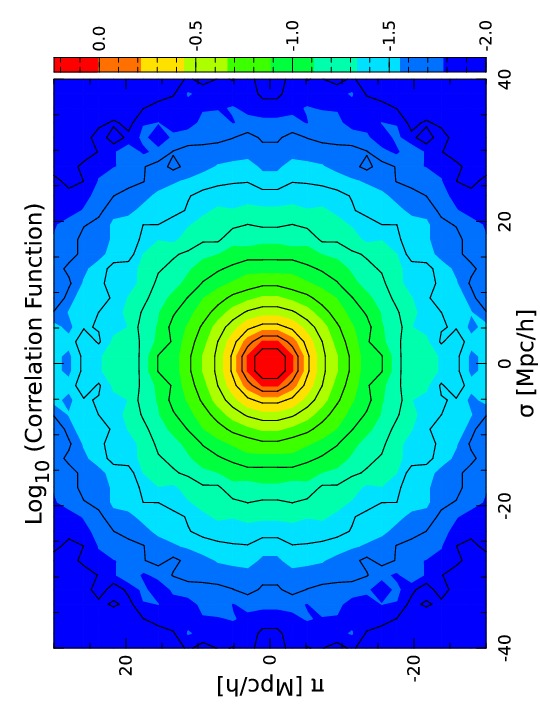} &
\includegraphics[angle=-90,width=0.5\linewidth]{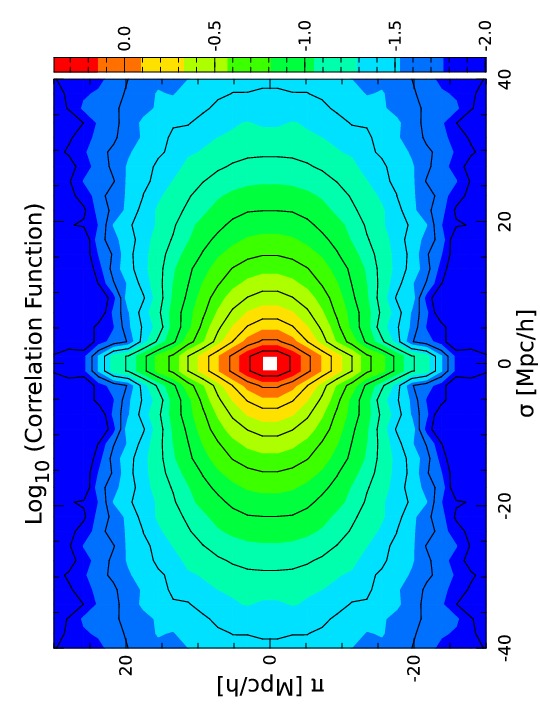} 
\end{tabular}
\caption{Measurements of the 2D real-space (left-hand panel) and
redshift-space (right-hand panel) correlation functions of dark
matter particles in the $z=0.6$ snapshot of the main GiggleZ
simulation.}
\label{fig:gigmatcf}
\end{figure*}

\begin{figure*}
\centering
\begin{tabular}{ccc}
\includegraphics[angle=-90,width=0.33\linewidth]{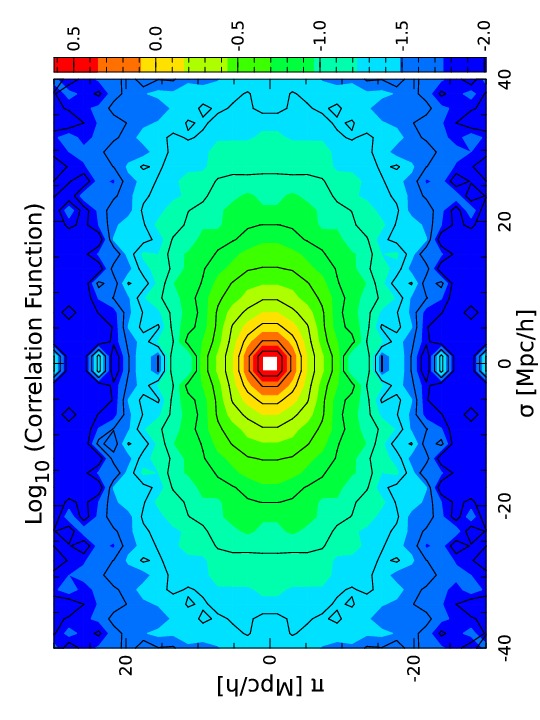} &
\includegraphics[angle=-90,width=0.33\linewidth]{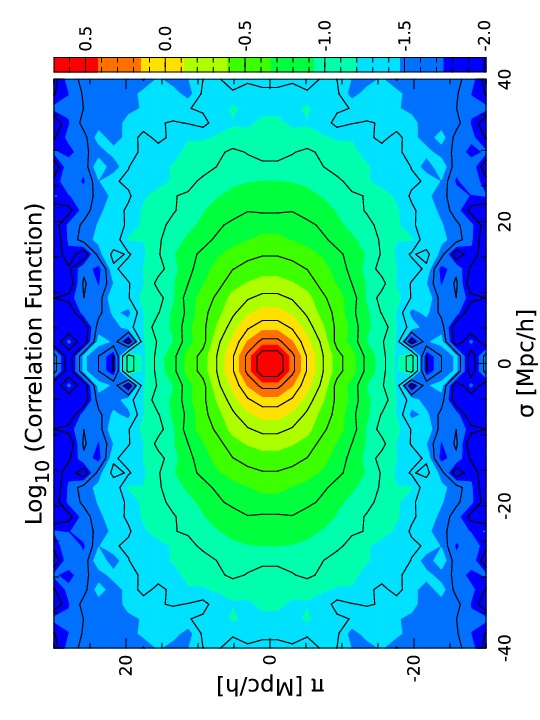} & 
\includegraphics[angle=-90,width=0.33\linewidth]{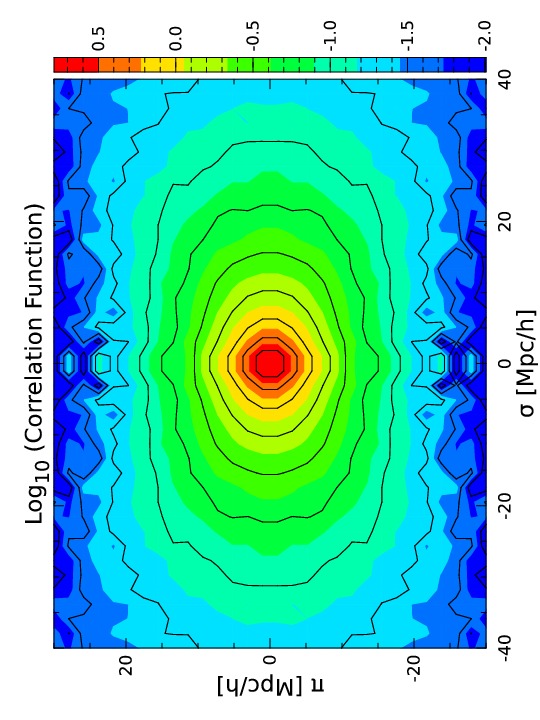} \\ 
\includegraphics[angle=-90,width=0.33\linewidth]{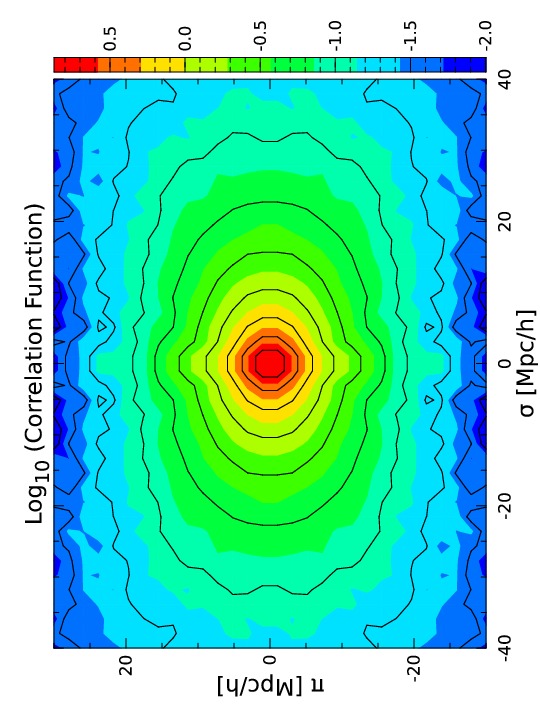} &
\includegraphics[angle=-90,width=0.33\linewidth]{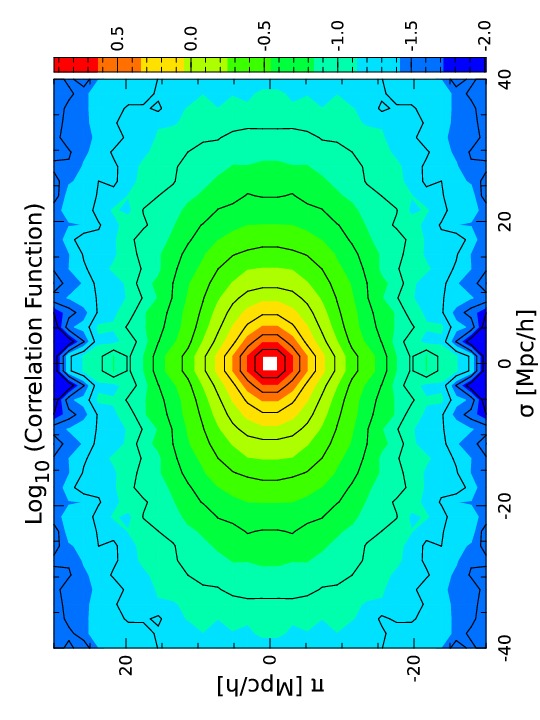} & 
\includegraphics[angle=-90,width=0.33\linewidth]{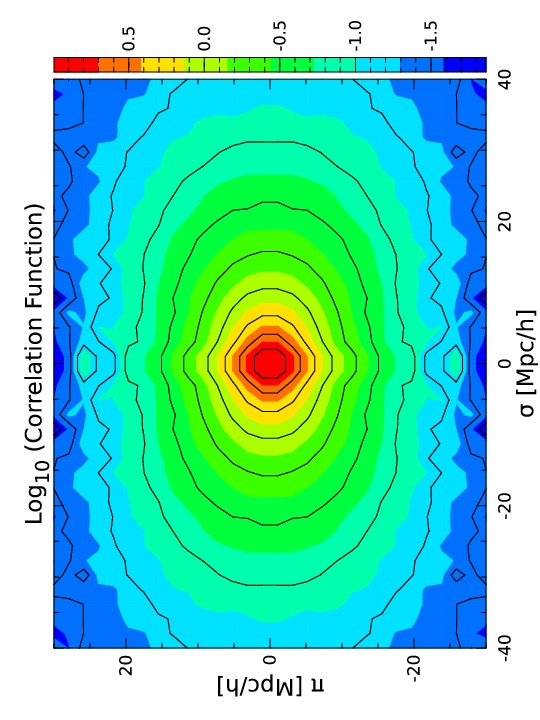}  
\end{tabular}
\caption{Measurements of the 2D redshift-space correlation function
for 6 different GiggleZ simulation halo-mass catalogues from low
mass (top left) to high mass (bottom right) in the $z=0.6$
snapshot.}
\label{fig:gighalocf}
\end{figure*}

We performed these measurements using simulation snapshots at
redshifts $z=0$ and $z=0.6$, in both real-space and redshift-space,
for the dark matter distribution and for the 6 different halo-mass
catalogues. We selected $z=0.6$ because this is the median redshift of
the WiggleZ Dark Energy Survey, and $z=0.0$ because here the
non-linearity in the clustering pattern that we are modelling will be
most significant.  We randomly sub-sampled the dark matter catalogue
for each snapshot to $10^6$ particles before performing the
correlation function estimation. The expected values for the growth
rate $f$ for these 2 snapshots are $0.49$ and $0.76$ respectively,
corresponding to the $\Lambda$CDM cosmological parameters used to
construct the simulations.  When generating redshift-space positions
we used a plane-parallel approximation, shifting co-ordinates along
one axis.  In Figure \ref{fig:gigmatcf} we plot the 2D real-space and
redshift-space correlation functions of the GiggleZ simulation dark
matter distribution for the $z=0.6$ snapshot.  In Figure
\ref{fig:gighalocf} we plot the corresponding 2D redshift-space
correlation functions of the 6 halo-mass catalogues.


\section{Modelling the redshift-space correlation function}
\label{sec:models}

\subsection{Constructing the model}

The fact that large-scale galaxy surveys observe redshifts, not
distances, implies that their clustering pattern is distorted by
galaxy peculiar velocities.  These peculiar velocities are generally
modelled in the correlation function by a combination of the two
effects which dominate in the large-scale and small-scale limit: the
large-scale coherent flow of galaxies into clusters and superclusters,
and the small-scale random motions of galaxies within virialized
structures.  The large-scale effects of coherent flows on the power
spectrum and correlation function can be described by the standard
treatment of \citet{Ka87} and \citet{H92} and the small-scale random
velocity distribution can be introduced by convolving with a function
$f(v)$, as summarized for example by \citet{Hk03}

We note some potential systematic errors in this approach, which we
discuss in turn in the remainder of this Section.  Firstly, the
isotropic real-space correlation function $\xi_r(s)$ must be modelled
reliably in order to extract the anisotropic signature.  Historically,
a power-law has been employed \citep{Hk03,Md03,CG09}.  However,
with increasing quality of data and simulations, a power-law has
become a bad approximation to the true non-linear clustering pattern.
We discuss some improvements below.  Secondly, the non-linear
behaviour of small-scale peculiar velocities is entirely modelled by
the single function $f(v)$.  However, in reality this function is
describing a scale-dependent combination of at least two physical
effects: the virialized motion of galaxies within haloes, and the
non-linearity in the coherent flows of galaxies which damp the
velocity power spectrum on quasi-linear scales
\citep{Sh96,Sl06}.  In detail, a single function is unlikely to
provide a good match in both regimes, and indeed more complex models
have been considered for matching small-scale data
\citep{CG09}.  Another possibility is to reduce the impact of
systematic modelling errors by excluding data at small scales from the
fit \citep{Hk03}, although this comes at the price of throwing
away a fraction of the data and in consequence obtaining a
statistically poorer measurement of the growth rate.

\subsection{Models for the real-space correlation function}

\subsubsection{Fitting formulae from CAMB and halofit}

For a given set of cosmological parameters, the matter power spectrum
(hence correlation function) at recombination can be numerically
calculated by solving Boltzmann's transport equation; a popular
publicly-available code which provides this solution is CAMB
\citep{LC99}.  The effect of the non-linear growth of structure
at an arbitrary redshift can be incorporated using the ``halofit''
recipe calibrated by N-body simulations \citep{Sm03}.  In this
model we can determine the galaxy correlation function by combining
this with a linear galaxy bias parameter $b$ as a simple
normalization.  We refer to the non-linear real-space galaxy
correlation function generated in this manner as the ``CAMB model'',
and by combining this real-space correlation function with the
redshift-space distortion parameters described above we can determine
the corresponding redshift-space correlation function:
\begin{eqnarray}
\xi_s(\sigma,\pi) = {\rm CAMBmodel} \left[ b, \beta, f(v) \right]
\end{eqnarray}
We explore below the dependence of the results on the fitted range of
scales.  We obtained the input CAMB power spectrum using the WMAP-7
best-fitting cosmological parameters \citep{La11} which are
consistent with the input parameters for the GiggleZ simulation.

\subsubsection{Power-law correlation function}

In previous studies the real-space galaxy correlation function at
small scales has often been modelled with a power-law form $\xi_r =
\left( \frac{r}{r_0} \right)^{-\gamma}$ and the full model for the
redshift-space correlation function can be written as
\begin{eqnarray}
\xi_s(\sigma,\pi) = {\rm PowerLawModel} \left[ \gamma, r_0, \beta, f(v) \right]
\end{eqnarray}

\subsubsection{Quadratic correlation function (\Qcfr )}

Improving data from both galaxy surveys and numerical dark matter
simulations has demonstrated that the real-space galaxy correlation
function deviates from a power-law at scales beyond $\sim 15 \,
h^{-1}$ Mpc \citep{Hk03}.  Thus in the case of high signal-to-noise
data, a power-law model produces a poor fit to the correlation
function.  This motivates our definition of a new empirical fitting
formula with greater flexibility than the CAMB and power-law models.
We introduce here the {\em quadratic correlation function (QCF) model}
\begin{eqnarray}
\xi_r(r)=\left(\frac{r}{r_0}\right)^{-\gamma + q \, {\rm log}_{10}\left(\frac{r}{r_0}\right)}
\end{eqnarray}
where $q$ is the additional quadratic parameter.  This is just a
simple quadratic equation in logarithmic space:
\begin{eqnarray}
y = a + b x + q x^2 \quad ; \quad x={\rm log}_{10}(r) \quad \& \quad y={\rm log}_{10}(\xi_r)
\end{eqnarray}
Our full model in this scenario is then
\begin{eqnarray}
\xi_s(\sigma,\pi) = {\rm QCFModel} \left[ \gamma, r_0, q_0, \beta, f(v) \right]
\end{eqnarray}
Although this model is not physically motivated, it produces an
impressive fit over the wide range of scales $1 - 50 \, h^{-1}$ Mpc to
both a suite of non-linear matter correlation functions generated by
CAMB, and (as we show below) to the real-space correlation function of
a range of halo-mass catalogues from the GiggleZ simulation.  The
flexibility of the QCF model enables us to achieve fits to the
redshift-space correlation function with lower systematic errors.  We
note that the QCF model, when applied to a halo correlation function,
can also accommodate a scale-dependent bias factor.

\begin{figure}
\centering
\includegraphics[width=75mm]{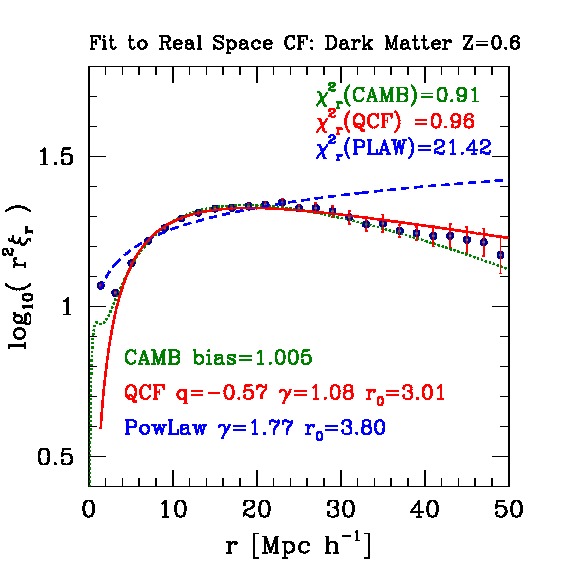}
\caption{Fit of the power-law, CAMB and QCF models (dashed, dotted 
  and continuous line-style respectively) to the real-space
  correlation function measured from the GiggleZ simulation
  dark-matter catalogue at redshift $z=0.6$.  The best-fitting values
  of the parameters and reduced chi-squared values $\chi^2_r$ are
  listed in the Figure. This plot shows that a Power Law model does
  not accurately describe the real-space correlation function, which
  causes systematic errors in the determination of the growth rate and
  galaxy bias parameters in this case. The CAMB and QCF models
  provide a better match to the measured correlation function.}
\label{fig:fitrealmatcf}
\end{figure}

Examples of these fits are given in Figure \ref{fig:fitrealmatcf}
and Table \ref{tab:hgcf}.  In Figure \ref{fig:fitrealmatcf} a
comparison is shown between the 1D real-space correlation function
measured from the GiggleZ simulation at redshift $z=0.6$, and fits of
the three real-space correlation function models defined above.  
In Table \ref{tab:hgcf} similar fits are shown for the 6
halo-mass catalogues.  We note that the power-law model always
produces a poor fit to the halo correlation function.  The CAMB model
describes the correlation function of low-mass haloes well, but breaks
down at higher mass due to the effects of scale-dependent bias.  The
QCF model produces a good fit to the data in all cases.

\subsection{Models for the pairwise velocity distribution}
\label{sec:stepwise}

In previous studies, two choices have been considered for the pairwise
velocity distribution $f(v)$ describing the random small-scale motions
of galaxies which result in the ``fingers-of-god'' observed in galaxy
redshift surveys: a Gaussian or exponential distribution
\citep{Rc98,LS98,Hk03}.  We write these functions as
\begin{eqnarray}
f_g(v)=\frac{1}{a\sqrt{2\pi}} exp \left( \frac{-v^2}{2a^2} \right)
\quad f_e(v)=\frac{1}{2a} exp \left( \frac{-|v|}{a} \right) ,
\end{eqnarray}
where $a$ is the standard deviation of both distributions.
Intuitively, a Gaussian distribution can be thought of as resulting
from virialized motions within a single dark matter halo, and an
exponential distribution can be thought of as a sum of these motions
across haloes of different mass \citep{Sh96}.  The exponential
distribution generally produces a better fit to data and simulations
\citep{HC98,LS98,Rc98,L02}.  For example, \citet{L02} demonstrated
that the Fourier transform of the pairwise velocity distribution of
three different galaxy surveys (LCRS, 2dFGRS and SDSS) is better fit
by a Lorentzian profile, implying an exponential distribution in
configuration space.

In this study we consider two variations.  Firstly we explore both
possibilities by fitting to our data the weighted combination
\begin{eqnarray}
 f_x(v) = x f_e(v) + (1-x) f_g(v)
\label{eqn:x}
\end{eqnarray}
thus adding a final parameter, $x$, to our correlation-function fits,
which we required to lie in the range $0 \le x \le 1$.

Secondly we considered expressing the pairwise velocity distribution
as a general stepwise function
\begin{eqnarray}
f(v) = a_i  \quad for \quad v_{i-1}<=|v|< v_i  \quad\quad  i=1,2,...N
\end{eqnarray}
for a number of intervals $N$ from $v_0=0$ up to some maximum pairwise
velocity $v_N=v_{max}$.  The distribution is normalized such that
$\int_{-\infty}^{\infty} f(v) dv = 1$, which implies that $2
\sum_{i=1}^{N} a_i (v_i - v_{i-1}) = 1$.  The quantities $N$ and
$v_{max}$ are set by hand (depending on the quality of the data) by
inspecting the solutions and requiring that $f(v)$ should be generally
positive and smoothly decreasing from $v=0$ to $v_{max}$.  For our
simulation dataset we typically obtain robust solutions using $N$ in the
range $6-9$ and $V_{max}$ between $1500-2500$ km s$^{-1}$.

For each combination of $\beta$ and the parameters describing the
real-space correlation function, we obtained the set of coefficients
$a_i$ which minimized the chi-squared statistic between model and data
by solving an $N \times N$ linear system of equations.  As explored in
more detail below, we obtain stable, physically sensible solutions
provided that we include correlation function measurements at small
transverse separations $\sigma$ in our fitted range (which have most
sensitivity to virialized motions).  Using this technique we can
determine the real underlying shape of the pairwise velocity
distribution, and test whether or not the exponential or Gaussian
models indeed provide a good description.  We give algebraic details
of this calculation in Appendix A.

\section{Model fits to the dark matter correlation function}
\label{sec:dmfit}

\subsection{Growth rate of the dark matter}

Firstly we explored how accurately the three real-space correlation
function models we have defined (CAMB, QCF, power-law) described the
real-space dark matter correlation function of the simulation, before
the addition of redshift-space distortions.  One example of these
results (for $z=0$) is displayed in Figure \ref{fig:fitrealmatcf}.
The minimum values of chi-squared, which are listed in Table
\ref{tab:dmcf}, demonstrate that the CAMB model provides the best fit
and a stable value of the galaxy bias $\sim 1$ for a variety of
fitting ranges, whilst the power-law model provides a poor fit to the
data.  The QCF model also yields a good fit for all scales $> 4 \,
h^{-1}$ Mpc.

\begin{table}
\begin{tabular*}{0.45\textwidth}{ @{\extracolsep{\fill}} |c|c|c|c|c|}
\hline
 CAMB & $r_{min}$ & CAMB & QCF & PLAW  \\
 Bias &  [$h^{-1}$ Mpc] & $\chi^2/dof$ & $\chi^2/dof$ & $\chi^2/dof$  \\ \hline
 1.006 & 0 &1.246 &  6.909 & 9.470 \\ \hline
 1.006 & 2 &1.234 &  1.687 & 9.425 \\ \hline
 1.006 & 4 &1.192 &  1.161 & 8.491 \\ \hline
 1.004 & 6 &1.019 &  0.931 & 6.438 \\ \hline
 1.000 & 8 &0.674 &  0.704 & 4.102 \\ \hline
 0.999 & 10 &0.633 &  0.669 & 2.559 \\ \hline
\end{tabular*}
\caption{This Table explores the effect of excluding the small-scale
  data bins from the fitting of the real-space dark matter correlation
  function. $r_{min}$ is the minimum value of the total separation ($r
  = \sqrt{\sigma^2 + \pi^2}$) for the data bins included in the fit.
  We note that the CAMB model provides a good fit to the real-space
  dark matter correlation function, even in the small-scale regime.
  The discrepancy in normalization between the fitted CAMB correlation
  function and the simulation is less than $1\%$.}
\label{tab:dmcf}
\end{table}

\begin{figure*}
\centering
\begin{tabular}{ccccc}
\includegraphics[angle=0,width=0.18\linewidth]{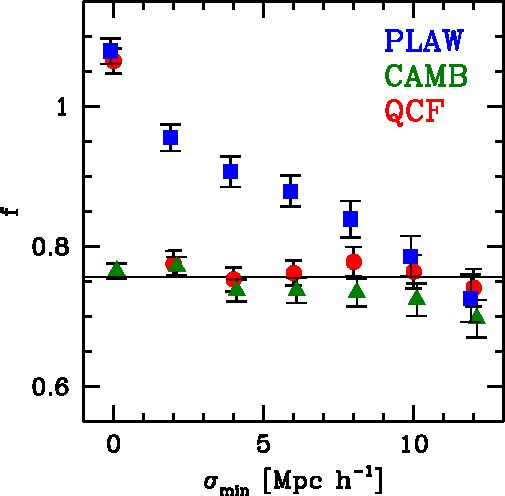} &
\includegraphics[angle=0,width=0.18\linewidth]{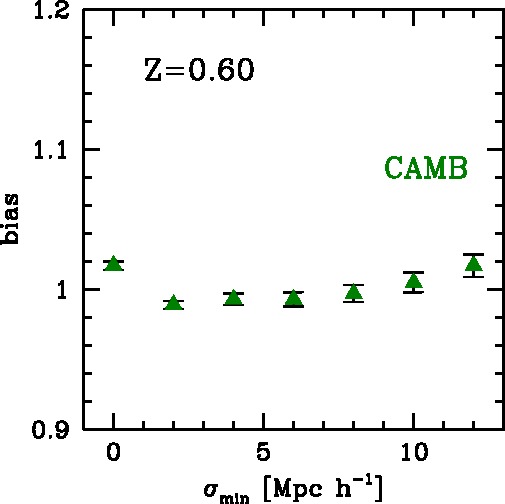} &
\includegraphics[angle=0,width=0.18\linewidth]{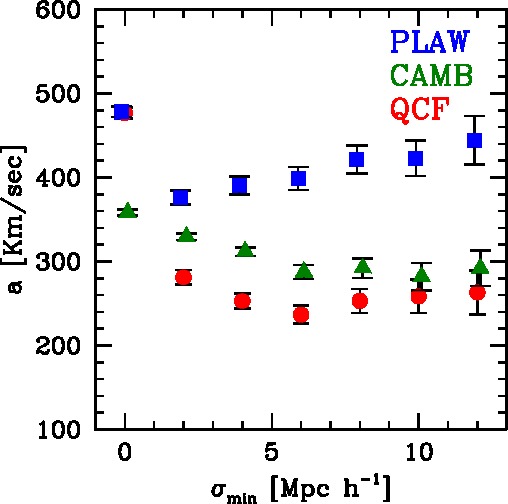} &
\includegraphics[angle=0,width=0.18\linewidth]{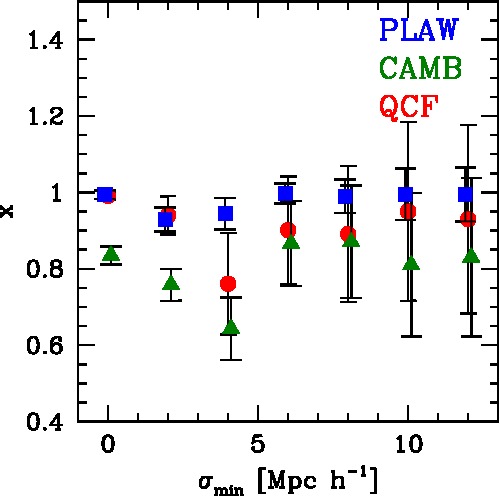} &
\includegraphics[angle=0,width=0.18\linewidth]{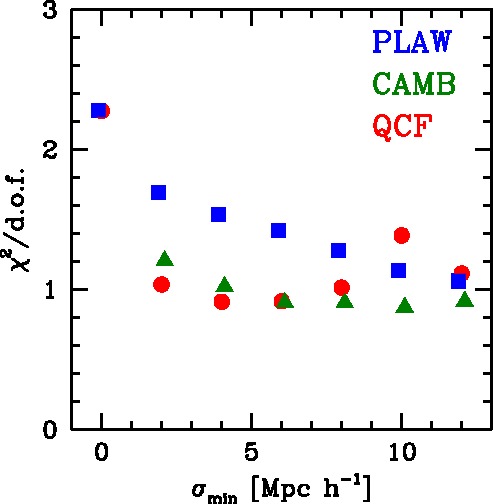} \\ 
\includegraphics[angle=0,width=0.18\linewidth]{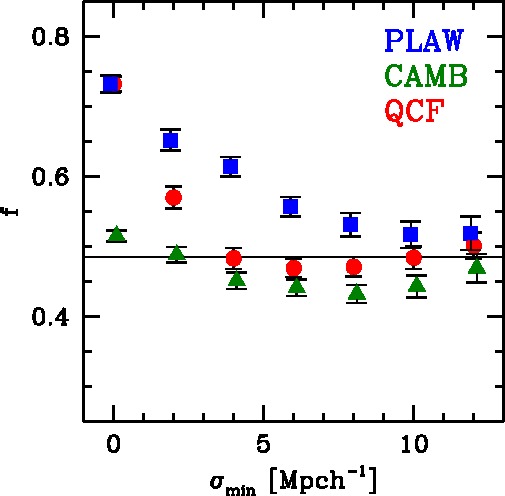} &
\includegraphics[angle=0,width=0.18\linewidth]{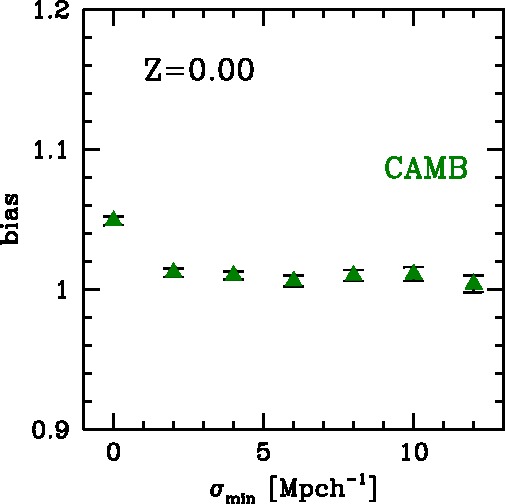} &
\includegraphics[angle=0,width=0.18\linewidth]{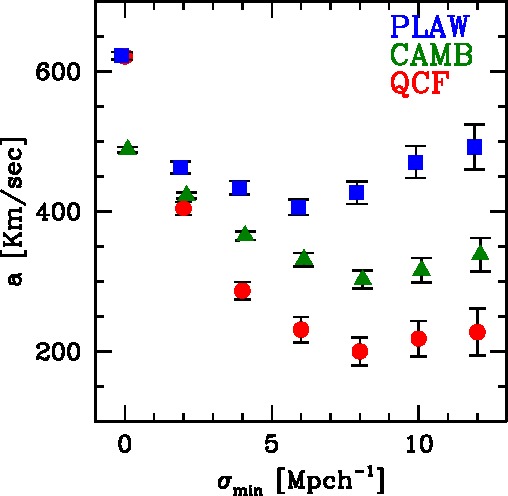} &
\includegraphics[angle=0,width=0.18\linewidth]{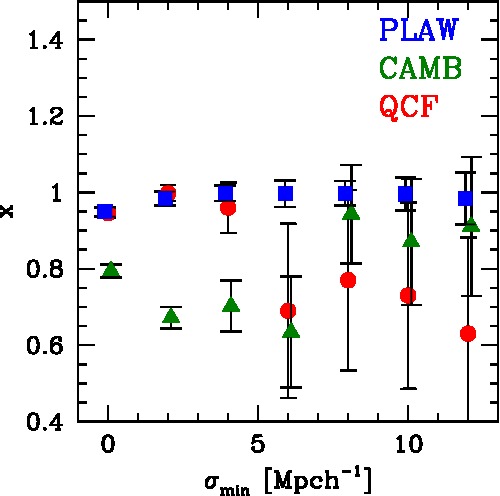} &
\includegraphics[angle=0,width=0.18\linewidth]{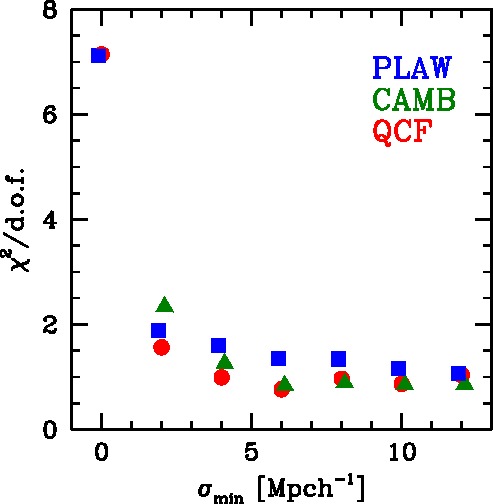}  
\end{tabular}
\caption{Fits for the growth rate, bias, the pairwise velocity
  dispersion, the parameter $x$ and the resulting reduced $\chi^2$,
  from the GiggleZ redshift-space 2D dark matter correlation function,
  for different values of the minimum transverse scale included in the
  fit, $\sigma_{min}$. Top: results from the $z=0.6$ snapshot of the
  simulation.  Bottom: results from the $z=0.0$ snapshot of the
  simulation.  Measurements are shown for three different real-space
  correlation function models.  The CAMB and QCF models (red circles
  and green triangles) show good agreement with the expected
  theoretical simulation growth rate (represented by the horizontal
  line) and rest of parameters, while the power-law model fit (blue
  squares) is strongly affected by systematics.The parameter $x$
  controls whether the pairwise velocity distribution is modelled as
  an exponential or a Gaussian, with the data clearly favoring an
  exponential form ($x \approx 1$).}
\label{fig:dmfit}
\end{figure*}

Next we included the effects of redshift-space distortions in the dark
matter correlation functions of the $z=0.0$ and $z=0.6$ snapshots, and
fit the clustering models described in Section \ref{sec:models}.
Figure \ref{fig:dmfit} displays how the measured parameters for the
two snapshots depend on the minimum transverse separation fitted,
$\sigma_{\rm min}$, for the three different real-space correlation
function models we are considering.  We fix the maximum transverse
separation fitted at $\sigma_{\rm max} = 40 \, h^{-1}$ Mpc.  We
performed the fits using a {\em Monte Carlo Markov Chain (MCMC)}
procedure, exploring the multi-dimensional space of variables of the
models and obtaining their joint and individual probability
distributions.  The horizontal lines indicate the input values of the
simulation, determined from its fiducial cosmological parameters.  We
find that the CAMB and QCF models are both able to recover the input
growth rate with low systematic error and reduced $\chi^2 \sim 1$, for
$\sigma_{\rm min} > 2 \, h^{-1}$ Mpc. On the other hand, the
assumption of a power-law model produces a significant systematic
error and bad fit.

\subsection{Pairwise velocity distribution of the dark matter}

In Figure \ref{fig:dmfit} we show the fitted values of the pairwise
velocity dispersion parameters $x$ and $a$, which are fit for each
model jointly with the other parameters.  The results strongly favor
an exponential rather than a Gaussian velocity distribution ($x
\approx 1$).  The systematic discrepancy between the values of $a$
fitted in the QCF and CAMB models indicate that there is some
cross-talk between $a$ and the shape of the real-space correlation
function on small scales.

We now compare these measurements with a direct determination of the
shape of the pairwise velocity distribution $f(v)$, which is possible
using our stepwise fitting method described in Section
\ref{sec:stepwise}.  This constitutes a further check for systematic
errors in the models which describe this distribution.  We fit the
stepwise distribution in 7 velocity bins up to a maximum velocity of
2000 km s$^{-1}$ (our results are not sensitive to these choices).  We
assumed a CAMB real-space matter correlation function in these fits.

\begin{figure*}
\centering
\begin{tabular}{ccc}
\includegraphics[angle=0,width=0.33\linewidth]{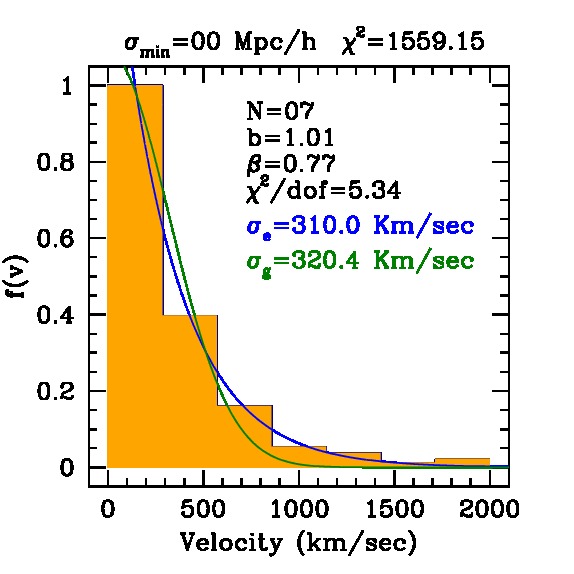} &
\includegraphics[angle=0,width=0.33\linewidth]{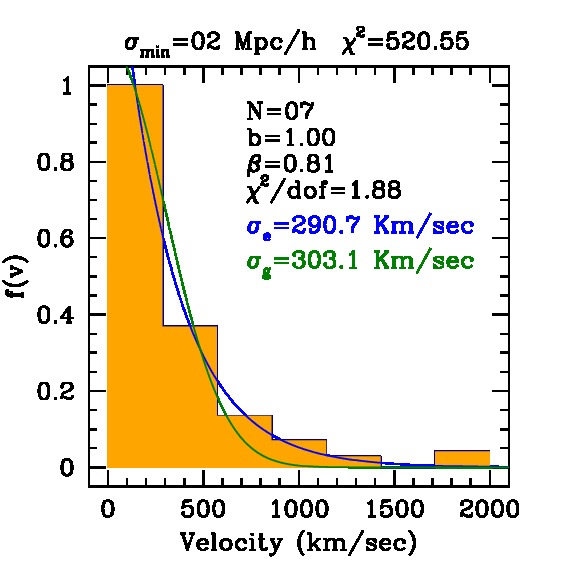} &
\includegraphics[angle=0,width=0.33\linewidth]{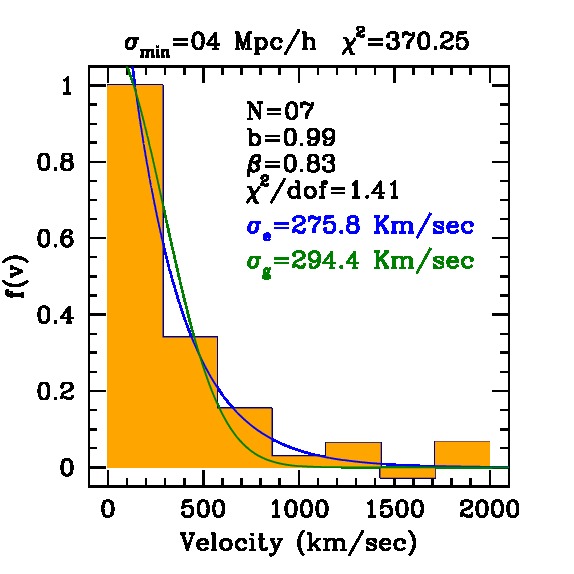} \\
\includegraphics[angle=0,width=0.33\linewidth]{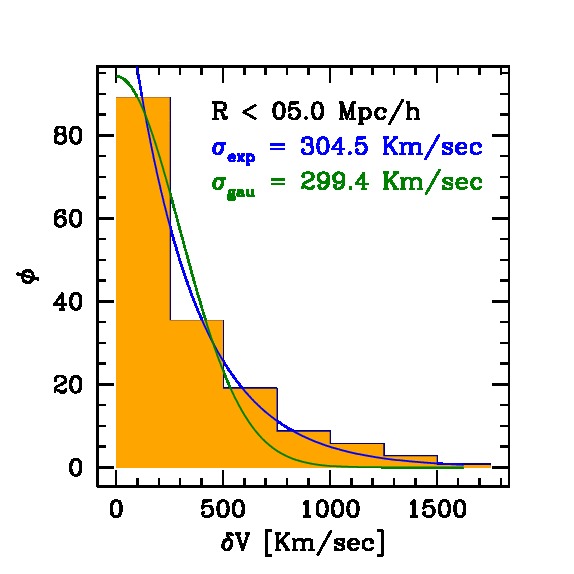} &
\includegraphics[angle=0,width=0.33\linewidth]{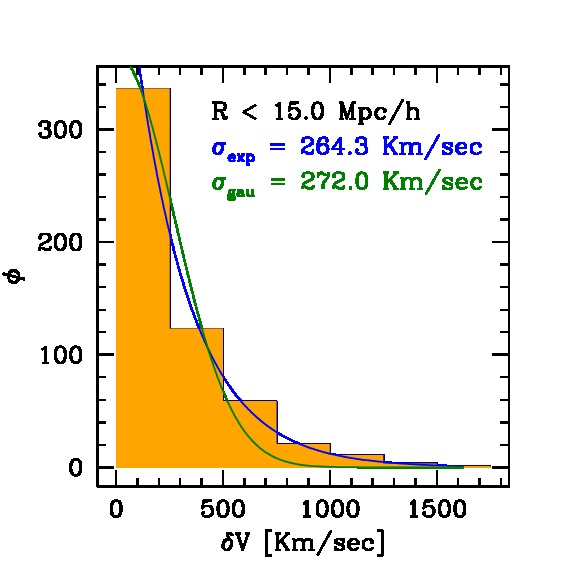} &
\includegraphics[angle=0,width=0.33\linewidth]{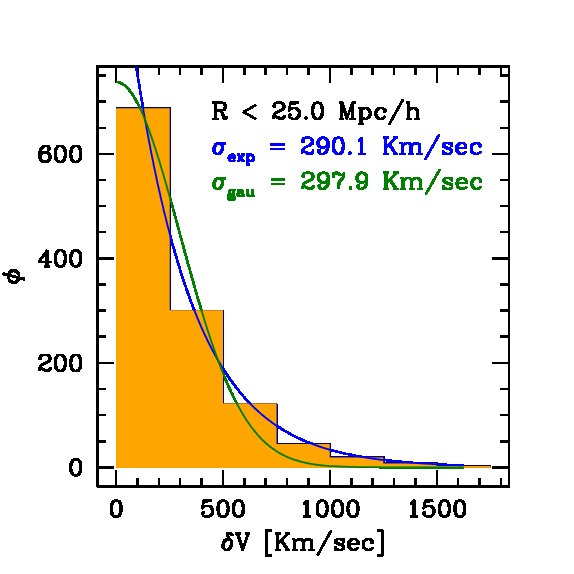}
\end{tabular}
\caption{Top row: The best-fitting stepwise functions for the pairwise
velocity distribution of the 2D dark matter redshift-space
correlation function, assuming a CAMB real-space model correlation
function.  The fits are repeated for 3 different values of the
minimum transverse separation $\sigma_{\rm min}$.  Bottom row:
direct measurements of the pairwise velocity distribution in the
dark matter catalogue, for different values of the maximum pair
separation $R$.  In both panels we fit Gaussian (green) and exponential (blue)
functions to the histograms. In the upper panel we restrict the
fitting to the positive and monotonic part of the data. From these
fits we get the value of the standard deviations. This plot
demonstrates that we are able to recover the pairwise velocity
distribution correctly using this method, and we find excellent
agreement between the measured velocity distribution function and
the exponential model.}
\label{fig:step_wise}
\end{figure*}

The results are displayed in Figure \ref{fig:step_wise}, comparing the
best-fitting stepwise distributions to both the best-fitting
exponential and Gaussian models, and to a direct measurement of the
pairwise velocity distribution from the catalogues.  In the top row,
the stepwise velocity distribution is fitted to the correlation
function of the dark matter subsample, for 3 different values of the
minimum value of $\sigma$ of the data bins included in the fit.  In
the bottom row the pairwise velocity distribution is directly measured
from the dark matter catalogues, this time varying the maximum total
separation ($R = \sqrt{\sigma^2+\pi^2}$) of the pairs considered for
this measurement.  We set a maximum value of $R$ for this calculation
because we only want to estimate the dispersion within a single bulk
flow, not between different bulk flows.  The distribution obtained
from the stepwise fitting process is in excellent agreement with its
direct measurement from the catalogue, demonstrating the capacity of
the stepwise approach to recover the pairwise velocity distribution
from the measured correlation function.  The distributions we obtain
are consistent with an exponential and not a Gaussian shape for
$f(v)$, and we recover the standard deviation of the pairwise velocity
distribution $a \sim 300$ km s$^{-1}$, independently of specifying a
model for $f(v)$.  This agreement between the model-independent
stepwise measurement and the assumed functional form increases our
confidence in the reliability of the exponential small-scale
redshift-space distortion model.


\section{Model fits to the halo correlation functions}
\label{sec:halofit}

We now turn our attention to fitting models to the halo correlation
functions, which represent galaxies in the simulation.  We explore how
accurately our empirical models can describe the halo clustering
in redshift-space.

\subsection{Redshift-space distortion parameters of the halo catalogues}

Table \ref{tab:hgcf} compares models and measurements in
real space, for the 6 halo-mass catalogues from the GiggleZ simulation
considered in this study.  The power-law model consistently fails to
match the measured correlation function.  The CAMB model produces a
good description of the clustering of low-mass haloes, but breaks down
for higher-mass catalogues due to scale-dependent halo bias.  The QCF
model is flexible enough to produce a good fit to the real-space
correlation function of all the halo catalogues.

\begin{table}
\begin{tabular*}{0.45\textwidth}{ @{\extracolsep{\fill}} |c|c|c|c|c|}
\hline
 Halo & CAMB &  CAMB & QCF & PLAW  \\
 Group& Bias &  $\chi^2/dof$ & $\chi^2/dof$ & $\chi^2/dof$  \\ \hline
 1 &1.069(005) & 1.403  & 1.370      & 2.461 \\ \hline
 2 &1.260(003) & 1.717  & 1.472      & 3.095 \\ \hline
 3 &1.417(003) & 1.494  & 1.475      & 4.753 \\ \hline
 4 &1.572(003) & 2.893  & 1.696      & 5.833 \\ \hline
 5 &1.707(003) & 1.815  & 1.368      & 4.810 \\ \hline
 6 &1.856(002) & 2.532  & 1.729      & 7.596 \\ \hline
\end{tabular*}
\caption{This Table compares models and measurements in
real space, for the 6 halo-mass catalogues from the GiggleZ simulation
considered in this study.  The power-law model consistently fails to
match the measured correlation function.  The CAMB model produces a
good description of the clustering of low-mass haloes, but breaks down
for higher-mass catalogues due to scale-dependent halo bias.  The QCF
model is flexible enough to produce a good fit to the real-space
correlation function of all the halo catalogues.}
\label{tab:hgcf}
\end{table}

\begin{figure*}
\centering
\includegraphics[width=150mm,angle=-90]{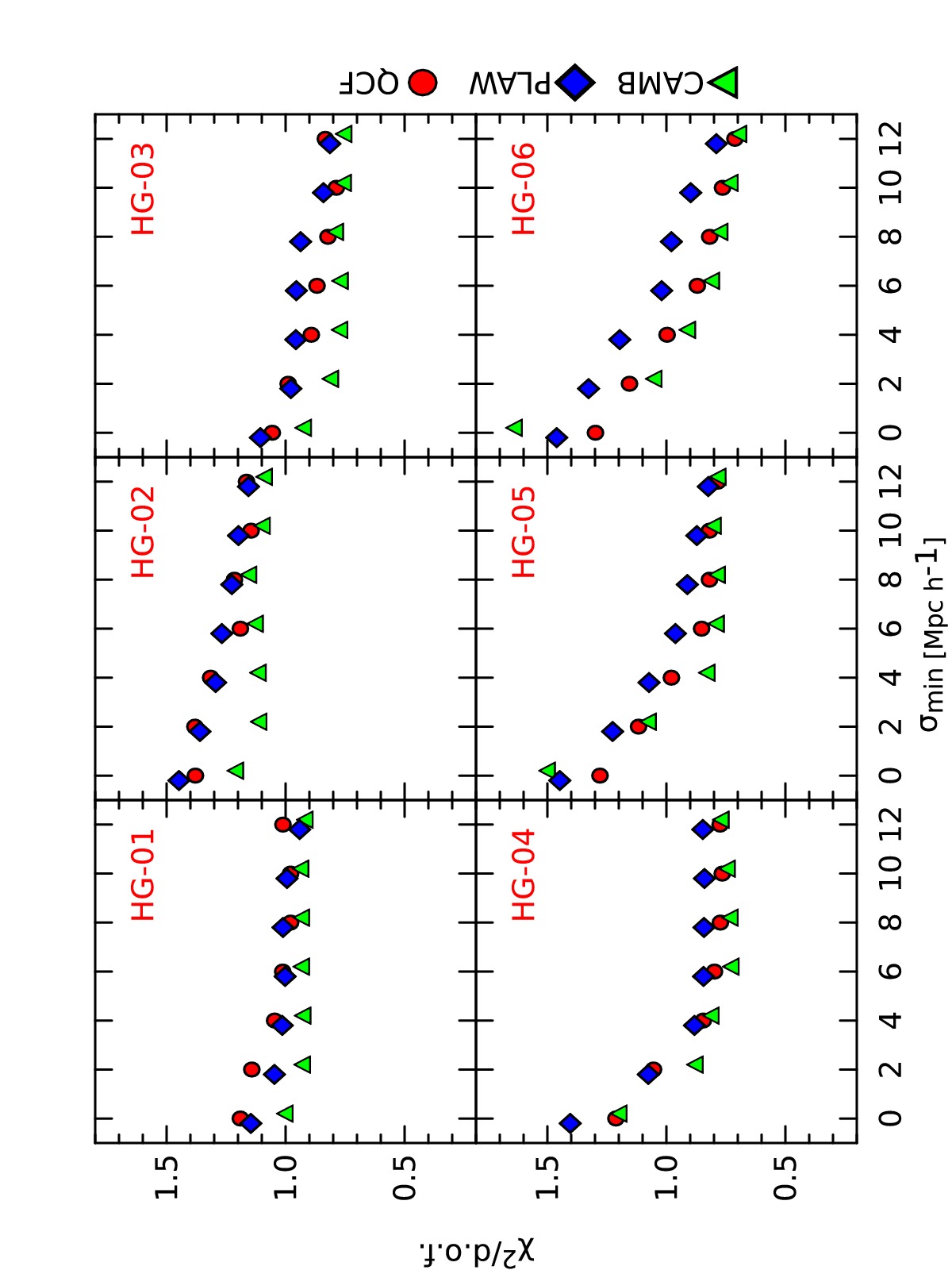}
\caption{The value of the reduced $\chi^2$ statistic of the
  best-fitting redshift-space distortion model as a function of the
  minimum value of $\sigma$ included in the fit.  Results are shown
  for each of the 6 halo-mass catalogues at redshift $z=0.6$, for the
  three different RSD models.}
\label{fig:chi}
\end{figure*}

We then fitted the redshift-space distortion models to the 2D
redshift-space correlation function measurements for these 6 halo-mass
catalogues.  In Figure \ref{fig:chi} we show the reduced $\chi^2$
statistic corresponding to fitting our three RSD models to the range
$\sigma_{min} < \sigma < 40 \, h^{-1}$ Mpc, $0 < \pi < 30 \, h^{-1}$
Mpc, as a function of the choice of the minimum value $\sigma_{min}$.
Our motivation for this analysis is that systematic errors in the
recovery of the redshift-space distortion parameter are likely to be
most serious at the lowest values of $\sigma$, where non-linear
velocity effects such as the ``fingers-of-god'' are most significant
(note however \citet{Bi12}, who show in a similar analysis that
increasing $\sigma_{min}$ does not make systematics completely
vanish).  We find that the power-law real-space model is a bad fit to
the data, whereas the CAMB and QCF models show mutual agreement and
better $\chi^2$ values.  The QCF model provides the best fit to the
clustering pattern, particularly for high-mass haloes.

The best-fit values for the pairwise velocity dispersion $a$ depend on
the real-space correlation function model in a similar way to the fits
to the dark matter catalogues.  For these halo catalogues the $x$
parameter, which was introduced to determine the shape of the pairwise
velocity distribution, does not favor the exponential function over
the Gaussian function as strongly as in the case of dark matter.  We
speculate that a halo catalogue with a narrow range of masses will
also possess a more uniform dispersion of pairwise velocities,
producing a closer match to a Gaussian function than before.  In order
to convert these results into growth rate measurements we require an
estimate of the galaxy bias, which we consider in the next Section.

\subsection{Bias factor of the halo catalogues}
\label{sec:bias}

Modelling the RSD in the 2-point correlation function gives us one of
the two quantities, $\beta$, that are necessary to deduce the growth
rate $f = \beta \, b$.  We now discuss and test possible methods to
measure the bias factor $b$, which describes the clustering of
galaxies (or dark matter haloes) relative to the underlying matter
distribution.  For example more massive haloes, which are sampled
preferentially from more clustered regions of the Universe, will
possess a higher bias factor $b$ and a correspondingly lower value of
$\beta$ (resulting in a less flattened large-scale redshift-space 2D
correlation function).  These more massive haloes are sites of early
galaxy formation which today can be observed as luminous red galaxies
\citep{Ei01,CG09}; less massive haloes preferentially host blue,
star-forming galaxies \citep{Md03,Blk09}.

The bias factor cannot be deduced directly from the measured 2-point
correlation function without some other assumption (such as the
underlying amplitude of the real-space matter correlation function),
although we note that such a determination may be possible using the
3-point correlation function \citep{Ve02,Ga05,Ma11}.  However, the use
of N-body simulation catalogues allows us to compare the bias
measurements resulting from fits of a real-space correlation function
model with those directly determined from the data via
\begin{eqnarray}
\label{eq:eqb}
b^2(r) = \xi_G(r)/\xi_{DM}(r) 
\end{eqnarray}
where $\xi_G(r)$ and $\xi_{DM}(r)$ are the real-space correlation
functions of the halo (galaxy) catalogue and dark matter distribution,
respectively. We assume hereafter bias is a constant in our fits,
which is a good approximation on large scales.

Table \ref{tab:biases} shows the consistency of the bias factors
resulting from fitting a CAMB model to the halo-mass catalogues, with
the direct measurements of the bias obtained by dividing the different
halo-mass correlation functions by the dark matter correlation
function and using equation \ref{eq:eqb}.  We divided the two
correlation functions in each different separation bin, and averaged
over each measurement as an independent estimate of the bias. We note
that the two independent measurements are broadly consistent, with the
main discrepancy appearing for small-scales due to systematic
non-linear effects.  This discrepancy can be ameliorated by
restricting the fitting range to exclude small scales, which are
polluted by the highest-amplitude systematic errors.

\begin{table}
\begin{tabular*}{0.45\textwidth}{ @{\extracolsep{\fill}} |c|c|c|c|c|}
\hline
Halo & Measured &  Fitted & Measured & Fitted  \\
Group& Bias     &  Bias   & Bias     & Bias  \\ \hline
& \multicolumn{2}{c|}{z=0.0}& \multicolumn{2}{c|}{z=0.6} \\ \hline
1 & 0.760(024) & 0.80(02) &	1.034(027) & 1.06(02) \\ \hline
2 & 0.825(027) & 0.85(02) &	1.197(031) & 1.23(02) \\ \hline
3 & 0.980(023) & 1.01(02) &	1.364(038) & 1.40(02) \\ \hline
4 & 1.079(022) & 1.12(02) &	1.532(029) & 1.56(02) \\ \hline
5 & 1.172(022) & 1.20(02) &	1.643(047) & 1.71(03) \\ \hline
6 & 1.248(023) & 1.27(02) &	1.792(053) & 1.87(03) \\ \hline
\end{tabular*}
\caption{Comparison between the fitted and measured bias factors for
  the 6 halo-mass groups, for snapshots at redshifts $z=0.0$ and
  $z=0.6$.  In general we recover consistent values for the bias, but
  we note a marginally-significant systematic discrepancy for
  higher-mass haloes.}
\label{tab:biases}
\end{table}

The broad agreement between the bias factors fitted by the CAMB model
to the 2D redshift-space correlation function, and those measured
directly from the 1D real-space correlation function of the
simulation, gives us confidence to use the CAMB bias factors to
measure the growth rate of our halo catalogues.

\subsection{Growth rate of the halo catalogues}

\begin{figure*}
\centering
\includegraphics[width=150mm,angle=-90]{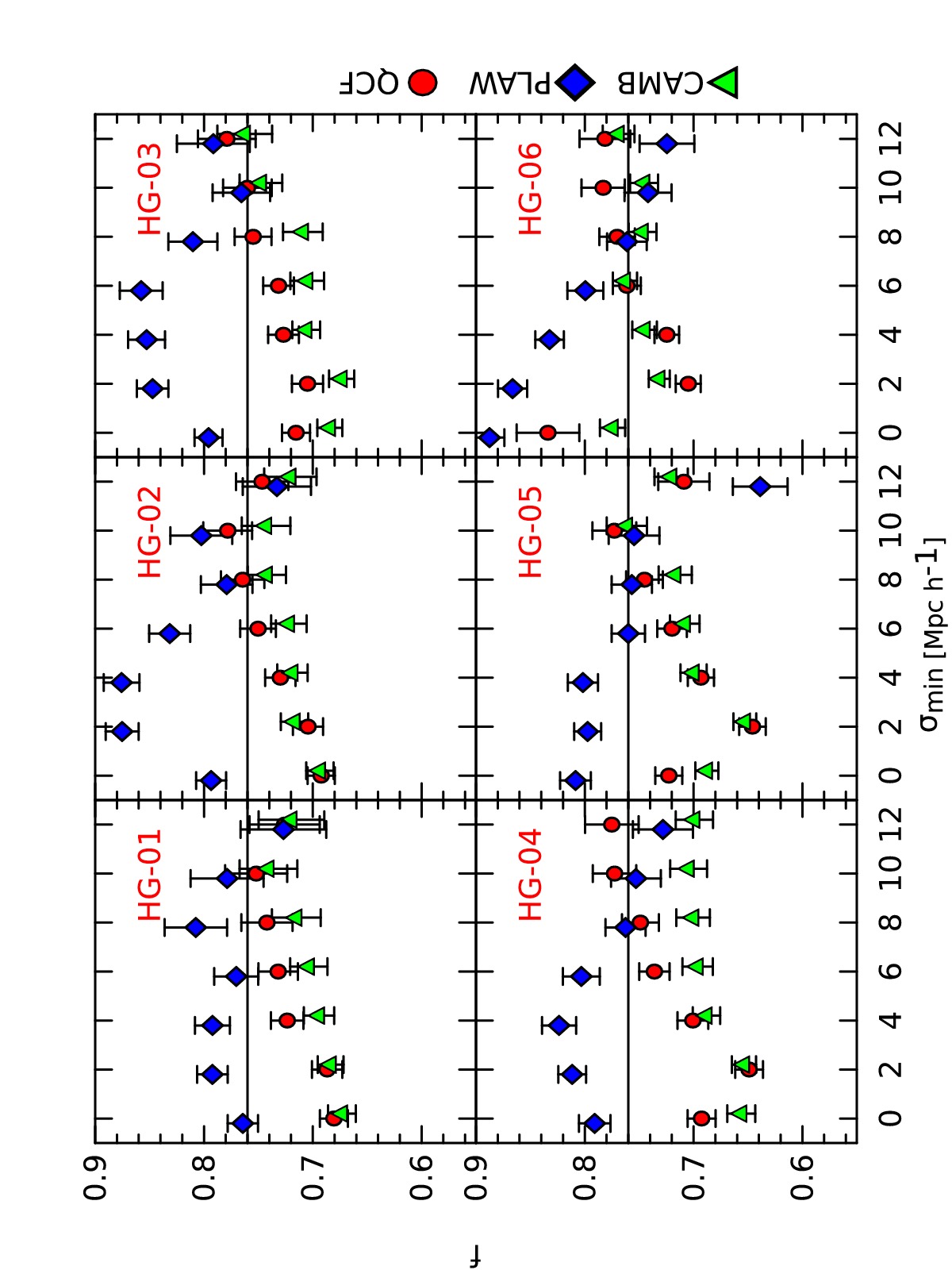}
\caption{The best-fitting value of the growth rate as a function of
  the minimum value of $\sigma$ included in the fit.  Results are
  shown for each of the 6 halo-mass catalogues for the $z=0.6$
  snapshot, for the three different RSD models. This figure shows we
  can recover the growth rate with minimal systematics when using QCF
  and CAMB models for the real-space correlation function.}
\label{fig:growth}
\end{figure*}

The resulting growth rate measurements for the $z=0.6$ simulation
snapshot, combining the separate determinations of the redshift-space
distortion parameter $\beta$ and the galaxy bias, are plotted in
Figure \ref{fig:growth}.  This figure illustrates the amplitude of the
systematic error in measuring the growth rate as a function of halo
mass, minimum transverse scale fitted, and model adopted for the
real-space correlation function.  All models contain a systematic
error $\Delta f \approx 0.05$ for $\sigma_{min} < 6 \, h^{-1}$ Mpc.
However, our fitting procedure, and the adoption of the QCF model,
allows us to recover accurate measurements when excluding the
small-scale region ($\sigma_{min} > 6\, h^{-1}$ Mpc).  Similar results
are obtained for the $z=0.0$ snapshot.  We note in general that the
model fits to halo catalogues described in this Section contain
somewhat higher systematic errors than the fits to the underlying dark
matter distribution.

\section{Conclusions}
\label{sec:conc}

We have measured the 2D correlation function of the dark matter and
halo-mass catalogues for 2 snapshots of the GiggleZ simulation at
redshifts $z=0.0$ and $z=0.6$, and fitted different clustering models
to recover the growth rate.  We carefully study the effect on the
results of the range of scales fitted and the halo mass, spanning halo
bias factors between unity (hosting emission-line galaxies) and
high-mass haloes traced by Luminous Red Galaxies.  We list our
conclusions as follows:

\begin{itemize}

\item The commonly-used power-law model for the real-space correlation
  function produces a poor fit to the clustering pattern and a
  systematic error in the resulting growth rate.  A real-space
  correlation function based on a non-linear CAMB model does better,
  but breaks down for high-mass haloes.  We introduce a new empirical
  model, the quadratic correlation function (QCF model, which has one
  more degree of freedom than a power-law), which provides a better
  description of the real-space correlation function (particularly for
  high-mass haloes which possess significant scale-dependent bias) and
  produces a measurement of the input growth rate with a lower
  ($5-10\%$) amplitude of systematic error.

\item We introduce a new technique which permits the measurement of
  the pairwise velocity distribution as a stepwise function from the
  redshift-space correlation function.  The pairwise velocity
  distribution measured directly from the simulation catalogues is
  consistent with this model, and matches closely to the exponential
  function expected from theoretical considerations.

\item We have quantified the amplitude of systematic error in the
  measured growth rate from our $1 \, h^{-3}$ Gpc$^3$ simulation as a
  function of halo mass, the model employed, and the minimum
  transverse scale $\sigma_{min}$ fitted.  We find that for
  $\sigma_{min} < 6 \, h^{-1}$ Mpc, the systematic measurement error
  using our procedure is $\Delta f \approx 0.05$.  The
  adoption of the QCF model allows us to recover accurate measurements
  of the growth rate from halo catalogues when excluding the small
  scale region ($\sigma_{min} > 6\, h^{-1}$ Mpc). This is consistent
  with the recent analysis of \citet{Bi12}.

\item We note that the halo correlation function contains a higher
  level of systematic modelling errors than the dark matter
  correlation function, due to scale-dependent galaxy bias.  Our
  modelling allows us to recover the growth rate from the dark matter
  particle catalogue with no detectable systematic error.

\end{itemize}

We conclude that N-body simulations are an essential tool for testing
and developing methods to measure the cosmic growth rate, and that our
empirical techniques should be useful for modelling correlation
functions measured in the latest large-volume galaxy redshift surveys.
A companion paper will represent the application of these methods to
the WiggleZ Dark Energy Survey dataset.

\section*{Acknowledgements}

We acknowledge financial support from the Australian Research Council
through Discovery Project grants funding the positions of GP and FM.
This research was supported by CAASTRO: http://caastro.org. CC thanks
Karl Glazebrook who suggested the stepwise approach, Eyal Kazin for
his useful comments on the previous draft of this work, and Mercedes
L\'opez-Morales (IEEC), Mercedes Moll\'a (CIEMAT), Ana Mar\'ia
Mart\'inez (UCLM) and Luis David Moya for their kind hospitality,
support and useful feedback.

\appendix

\section{Stepwise Velocity Distribution Method}

As an alternative to assuming a Gaussian or exponential pairwise
velocity distribution, we explored fitting a stepwise function
($f(v)=a_k$ in a range of velocities) to the data, together with the
other model parameters.  In order to avoid introducing a large number
of additional free parameters, we have developed a method which allows
these fits to be performed in a relatively simple and fast way.

The standard expressions for the correlation function multipoles in
the large-scale limit \citep{Ka87,H92} are
\begin{eqnarray}
\label{eq:eq-gen}
\xi_0(r) &=& \left(1 +\frac{2\beta}{3}+\frac{\beta^2}{5}\right) \xi_r(r) \\
\xi_2(r) &=& \left(\frac{4\beta}{3}+\frac{4\beta^2}{7}\right)\left[\xi_r(r)-\dot
{\xi}_r(r) \right] \\
\xi_4(r) &=& \left(\frac{8\beta^2}{35}\right)  \left[\xi_r(r)+\frac{5}{2}\dot{\xi}_r(r)-\frac{7}{2}\ddot{\xi}_r(r) \right] 
\end{eqnarray}
where
\begin{eqnarray}
\dot{\xi}_r(r) &=& \frac{3}{r^3}\int_0^r{\xi_r(s) s^2 ds} \\
\ddot{\xi}_r(r) &=& \frac{5}{r^5}\int_0^r{\xi_r(s) s^4 ds} 
\end{eqnarray}
Starting from the real-space correlation function, such as obtained
from a CAMB matter power spectrum, we re-write equation
\ref{eq:eq-gen} in the form:
\begin{eqnarray}
\xi_0(r) &=& C_0(\beta) J_0(\sigma,\pi) \nonumber \\
\xi_2(r) &=& C_2(\beta) J_2(\sigma,\pi) \nonumber \\
\xi_4(r) &=& C_4(\beta) J_4(\sigma,\pi) 
\end{eqnarray}
where the form of the functions $C_i$ and $J_i$ can be deduced by
analogy with equation \ref{eq:eq-gen}, and we solve the integrals
numerically.  Now replacing the velocity distribution $f(v)$ by a
stepwise function $a_k$ we obtain:
\begin{eqnarray}
\xi(\sigma,\pi)=\int_{-\infty}^{\infty}{\xi'(\sigma,\pi')f(v)dv}= \nonumber \\
\sum_{k=1}^N{a_k\left[\int_{-a_k}^{-a_{k-1}}{\xi'(\sigma,\pi')dv}+\int_{a_{k-1}}^{a_k}{\xi'(\sigma,\pi')dv}\right]}
\end{eqnarray}
where we separate the negative and positive parts of the integral over $v$
because of the loss of symmetry implied by the relation
\begin{eqnarray}
\pi' = \pi + \frac{v(z+1)}{H(z)}
\end{eqnarray}
for the convolution with the pairwise velocity distribution.  We now
numerically calculate the following terms:
\begin{eqnarray}
Q_0(\sigma,\pi,k)=\int_{a_{k-1}}^{a_k}{J_0(\sigma,\pi) P_0 (\mu)dv} \nonumber \\
Q_2(\sigma,\pi,k)=\int_{a_{k-1}}^{a_k}{J_2(\sigma,\pi) P_2 (\mu)dv} \nonumber \\
Q_4(\sigma,\pi,k)=\int_{a_{k-1}}^{a_k}{J_4(\sigma,\pi) P_4 (\mu)dv}
\end{eqnarray}
where $P_i(\mu)$ are the Legendre polynomials in terms of $\mu$, the
cosine of the angle between the bin position and the line-of-sight
axis $\pi$.  We also define analogous functions $Q'$ integrating over
the negative part of the $v$ axis.  Now we combine these functions for
a given value of $\beta$ to create a series of stepwise models
$\xi(\sigma,\pi)=\sum{a_k F(\beta)_{\sigma,\pi,k}}$, where the final
$F_{\sigma,\pi,k}$ terms correspond to:
\begin{eqnarray}
F(\beta)_{\sigma,\pi,k}=C_0(\beta) \left[ Q_0(\sigma,\pi,k) + Q_0'(\sigma,\pi,k)\right] \nonumber \\
       +C_2(\beta) \left[ Q_2(\sigma,\pi,k) + Q_2'(\sigma,\pi,k)\right] \nonumber \\
       +C_4(\beta) \left[ Q_4(\sigma,\pi,k) + Q_4'(\sigma,\pi,k)\right] 
\end{eqnarray}
Now we have a model which is linearly dependent on $a_k$, which
constitutes an $N\times N$ linear system of equations.  Our $\chi^2$
equation is:
\begin{align}
\chi^2(\beta)=\sum_{i,j}\left[\left(\sum_k{a_k F(\beta)_{i,k}}-D_i\right) C^{-1}_{i,j} \left(\sum_k{a_k F(\beta)_{j,k}}-D_j\right)\right]
\end{align}
where $D$ is the data array, $C$ is the covariance matrix,
$i$ and $j$ represent bins in the correlation function data, 
and $k$ is the index for the stepwise velocity distribution.

The $n^{th}$ equation of the $N\times N$ linear system is:
\begin{align}
\sum{a_k  \sum_{i,j} C^{-1}_{i,j} \left[ F(\beta)_{i,k}F(\beta)_{j,n}+F(\beta)_{j,k}F(\beta)_{i,n}\right]} = \nonumber \\ 
\sum_{i,j} C^{-1}_{i,j} \left[ D_i F(\beta)_{j,n} +D_j F(\beta)_{i,n} \right]
\end{align}
The linear system may be solved by conventional methods and the
parameter space in $\beta$ can be quickly explored in search of the
minimum $\chi^2$.  The normalization of the stepwise function gives in
this case the factor between the fitted galaxy correlation function
and the CAMB matter correlation function, i.e.\ the bias $b^2$.  If
the real-space correlation function is a power-law, the first set of
numerical integrations may be replaced by analytical expressions, with
the clustering length $r_0$ of the power-law absorbed into the
normalization factor.

\bibliography{mybib}
\bibliographystyle{mn2e}
\end{document}